# The Early Statistical Years: 1947–1967 A Conversation with Howard Raiffa

**Stephen E. Fienberg**

*Abstract.* Howard Raiffa earned his bachelor's degree in mathematics, his master's degree in statistics and his Ph.D. in mathematics at the University of Michigan. Since 1957, Raiffa has been a member of the faculty at Harvard University, where he is now the Frank P. Ramsey Chair in Managerial Economics (Emeritus) in the Graduate School of Business Administration and the Kennedy School of Government. A pioneer in the creation of the field known as decision analysis, his research interests span statistical decision theory, game theory, behavioral decision theory, risk analysis and negotiation analysis. Raiffa has supervised more than 90 doctoral dissertations and written 11 books. His new book is *Negotiation Analysis: The Science and Art of Collaborative Decision Making.* Another book, *Smart Choices*, co-authored with his former doctoral students John Hammond and Ralph Keeney, was the CPR (formerly known as the Center for Public Resources) Institute for Dispute Resolution Book of the Year in 1998. Raiffa helped to create the International Institute for Applied Systems Analysis and he later became its first Director, serving in that capacity from 1972 to 1975. His many honors and awards include the Distinguished Contribution Award from the Society of Risk Analysis; the Frank P. Ramsey Medal for outstanding contributions to the field of decision analysis from the Operations Research Society of America; and the Melamed Prize from the University of Chicago Business School for *The Art and Science of Negotiation.* He earned a Gold Medal from the International Association for Conflict Management and a Lifetime Achievement Award from the CPR Institute for Dispute Resolution. He holds honorary doctor's degrees from Carnegie Mellon University, the University of Michigan, Northwestern University, Ben Gurion University of the Negev and Harvard University. The latter was awarded in 2002.

*Stephen E. Fienberg is Maurice Falk University Professor of Statistics and Social Science, Department of Statistics and Machine Learning Department, Carnegie Mellon University, Pittsburgh, Pennsylvania 15213-3890, USA (e-mail: Fienberg@stat.cmu.edu).*

This conversation took place as part of an informal seminar in the Department of Statistics at Carnegie Mellon University on April 3, 2000, preceding by a day a seminar at which Howard Raiffa received the 1999 Dickson Prize in Science from the University. Others present and participating in the discussion included William Eddy, Rob Kass, Jay Kadane and Raiffa's wife of 55 years, Estelle. The topic of Howard's presentation at the Dickson ceremony was: "The Analytical Roots of a Decision Scientist." For the Department of Statistics he elaborated on the years 1947–1967.

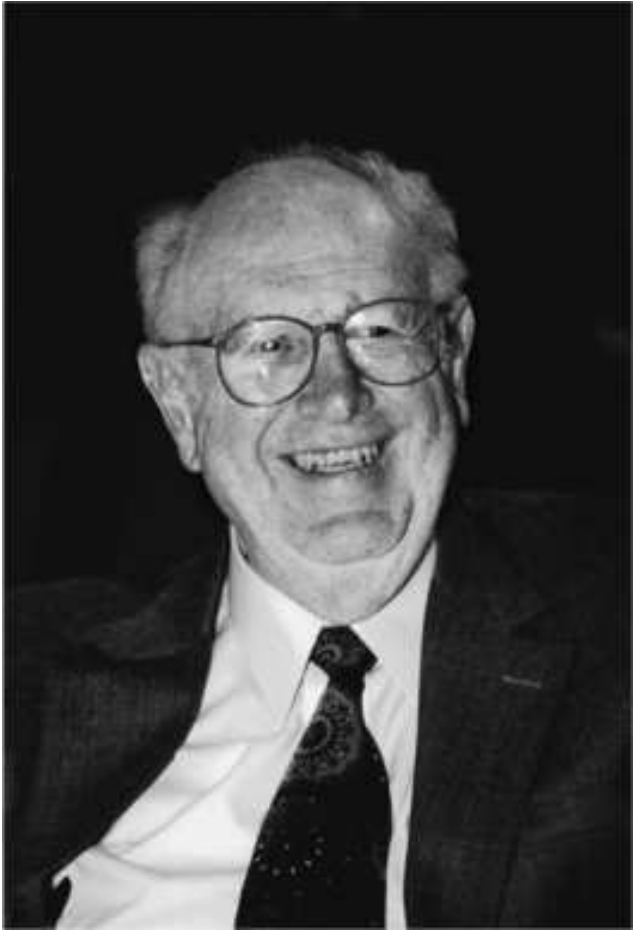

Fig. 1. *Howard Raiffa following the interview. April, 2000.*

**Fienberg:** In 1964 I arrived as a graduate student at Harvard and in my first class on statistical inference, a faculty member, whose name I will not mention, began teaching inference from a Neyman–Pearson perspective, that is, hypothesis testing and confidence intervals. In fact, we studied Erich Lehmann's book on hypothesis testing. It was clear to me that this wasn't the way I wanted to think about statistics. I asked around the department about what alternatives were available to me and someone said: "On Monday afternoons they have a seminar at the business school across the Charles River." So I just showed up one Monday afternoon. It was one of the most wonderful experiences of my graduate career. Although I have no memory of who was speaking that first afternoon, I recall that it was the most animated and heated discussion I had engaged in at any point in my career up to that point. It turned out that the seminar and the heated discussion were replicated every Monday afternoon. One of the leaders of that Monday afternoon seminar was Howard Raiffa.

**Raiffa:** That was called the Decision Under Uncertainty seminar, the DUU seminar, and it was one of the exciting parts of my life as well. And it went on for four years, from 1961 to 1964.

I ran that seminar with my colleague Robert Schlaifer. Between the weekly sessions of the seminar, a few of us exchanged a flood of memos commenting on what was discussed and what should have been discussed. Although there was a demand for air time, we never scheduled starting a new topic until we felt that we had completed the train of research thinking on the table. Half-baked ideas were given priority.

**Fienberg:** I was lucky to be able to attend for a couple of them! I brought along with me a couple of slices of statistical history this afternoon: *Applied Statistical Decision Theory* (Raiffa and Schlaifer, 1961) and *Introduction to Statistical Decision Theory* (Pratt, Raiffa and Schlaifer, 1995). They come from the same period as the DUU seminar and both books had a tremendous impact on the Bayesian revival of the 1960s. I'm hoping we will have a chance to talk about them in a few minutes, but let's go back a little further in time and start with how you became a statistician.

**Raiffa:** Let me start at a decision node. I was a First Lieutenant in the Air Force in charge of a radar-blind-landing system at Tachikawa Airfield near Tokyo—ground-controlled approach it was called—and I was due for a routine discharge after completing 44 months in the armed services. Should I continue in Japan acting in the same extraordinarily exciting capacity as a civilian, earning oodles of money, or should I return to the States to complete my bachelor's degree, and if so, where, and to study what? I was 22. I thought of becoming an engineer building upon my practical experience with radar. I learned from an army buddy of a field I had never heard of: actuarial mathematics. Merit counted in the actuarial profession since budding actuaries had to pass nine competitive exams. The place to go to prepare for the first three of them was University of Michigan.

**Fienberg:** Howard, I thought you went to City College.

**Raiffa:** Before going into the army, I was more interested in athletics than scholarship. I was the captain of my high school basketball team in New York City—a high school of 14,000 students—at a time when basketball was the rage in NYC. I was a mediocre B student in my freshman and sophomore years at City College of New York (CCNY), the college of choice for poor and middle-income students in New York—I was on the poor side. Four years later, when I returned to college at the University of Michigan, I became a superb student. That surprised me. I also was a married man determined to become employable after getting my bachelors degree. For a while I thought the students at CCNY were just brighter than those in Michigan, but on reflection I didn't like that explanation. Something changed within me. I was shooting for straight A's and getting them.

**Fienberg:** We could statistically investigate this change if we had the right sort of data. But Howard, did you ever take those actuarial exams and how come you stayed on and got your master's degree in statistics?

**Raiffa:** I got my bachelor's in Actuarial Math and passed the first three exams and was on my way to becoming an actuary preparing for the fourth exam when I decided that I really wanted to study something more cerebral—something more theoretical.

**Fienberg:** Why statistics?

**Raiffa:** The second actuarial exam was on probability theory and we actuarial students took a course that delighted me. It used Whitworth's *Choice and Chance*, full of intriguing puzzlers on combinations and permutations. Brain teasers galore. The book was short on theory and long on problems and I loved it, and I was good at it. I just couldn't put it down. I was good enough that my professors encouraged me to go on and get a master's degree. So I took a master's degree in statistics. Frankly, I was also disappointed in that as well because, at that time, the statistics program in the mathematics department was short on theory and depth and long on computational manipulations. I remember having to invert an $8 \times 8$ matrix. People are laughing because now you enter an $8 \times 8$ matrix in your computer and poof, out comes the answer. But in those days we had a mechanical Marchant calculator and I hated it. Even though I won a prize for speed, it did me in.

**Fienberg:** So how come you stayed on for a doctorate?

**Raiffa:** Along with the courses I took in statistics, I also took for cultural curiosity a course in the foundations of mathematics by a Professor Copeland who later became one of my mentors. Copeland taught the course using the R. L. Moore pedagogical style. Are you familiar with this approach, Steve?

**Fienberg:** Yes, I am. But please tell us a little about the R. L. Moore pedagogy anyway. I understand that Jimmy Savage also became enamored with studying mathematics after being exposed to the Moore approach.

**Raiffa:** I never knew this until I attended a memorial service for Jimmy. Well, I was similarly affected. I was studying for a master's degree in statistics; going for a doctorate was not an alternative in my personal decision space—it just wasn't, but it should have been if I had only known more about making smart choices.

Copeland's first assignment was weird: "Here are some seemingly unrelated mathematical curiosities. Think about them. Try to make some conjectures about them. Try to prove your conjectures. Try to discover something of interest to talk about." I drew a blank and I came to class with nothing to contribute. So did twenty other students.

At the beginning of class on that first day the instructor asked, "Does anybody have any contributions to make?" We sat and sat and sat and ten minutes went by and he said, "Class dismissed." He added, "The same assignment tomorrow." The following day, he started the class: "Anybody have anything to say?" Finally, someone raised their hand and asked a question. The course was pure R. L. Moore. No books were used, absolutely no books. It was taboo to look at the literature because you might find hints. You should act as if you were a mathematician in the 17th century trying to prove something new. No matter that we were "discovering" well-known results; it was new to us. We students did not study mathematics; we did mathematics. The R. L. Moore method of teaching turned me on. I knew then that I wanted to become a mathematician because it was so much fun and, to my surprise, I found out that I was pretty good at it.

So I became a student of pure mathematics and I was deliriously happy. Thanks to my wife, who by that time became an elementary school teacher and could support me in a manner that I grew accustomed to, we went from abject poverty to solid riches.

In the year I studied statistics, I don't think I heard the word Bayes. As a way of inference it was nonexistent; inference was all strictly done from the Neyman–Pearson perspective. And the version of Neyman–Pearson statistics I was exposed to wasn't very the-oretical as well. They talked about tests of significance but they really didn't talk about the power of the tests.

**Eddy:** Who were your professors?

**Raiffa:** Paul Dwyer and Cecil C. Craig.

**Fienberg:** And they were all stalwarts of the Institute of Mathematical Statistics in its first couple of decades. So you weren't turned on by the kind of statistics they did?

**Raiffa:** They were good mathematicians and good statisticians; but to them, statistics meant something quite restrictive. Dwyer was computational and Craig did multivariate sampling theory. They were very good at what they did, but there was not a decision bone in their bodies.

**Fienberg:** So you are now studying pure mathematics. How did your interest in game theory start?

**Raiffa:** For ten hours a week I worked as a research assistant on an Office of Naval Research (ONR)-sponsored research program administered jointly by the Mathematics Department and the School of Engineering. My role was to attend national meetings and listen to applied problems that people were talking about that related to our ONR project and to try to formulate interesting mathematical problems for my more mathematical-oriented colleagues to work on. I became quite adept at taking ill-formed situations and translating them into mathematical problems that other people could work on, including myself.

Because submarine warfare was a hot topic, I read von Neumann's and Morgenstern's book on Game Theory—or at least parts of it. Jerry Thompson, a fellow student, and I developed a paper on how to solve two-person zero-sum games and we found out to our delight that this same algorithm could solve linear programming problems. At that time we didn't know anything about the simplex method, so for maybe two weeks we had the best algorithm for solving linear programming problems. Incidentally, Jerry has spent his distinguished academic career here at Carnegie Mellon University.

At that time there was no well-developed theory of the simplest two-person, non-zero-sum games—there still isn't. I (and probably hundreds of others unknown to me) investigated the many qualitatively different bi-matrix games having two strategies for each player. I, naturally, became intrigued with a game now known as the Prisoner's Dilemma game. I know you are familiar with this game. The point is that each player has a dominant strategy so the optimal thing for each to do is to choose this strategy. But the rub is that if each player plays wisely, they each get miserable payoffs. The paradox is that two wise players do worse than two dumb players. Still, in a single-shot situation, each player should choose wisely. Rational individual choice leads to group inefficiencies. The anomaly lies in the structure of the game.

When the game is repeated for a pre-specified number of iterations, equilibrium analysis specifies that each player should use his or her dominating strategy at each trial with the result that each does miserably trial after trial. The game presents a social pathology. If the players had pre-play communication and could make binding agreements, they would agree to cooperate at each trial by taking the myopically dominated strategy. In the finitely repeated game, double crossing at each trial (i.e., choosing the non-cooperative strategy at each trial) is the best retort if the other guy acts that way. But double crossing at each trial is not the best retort against someone who is not playing the double cross strategy at every trial. In the laboratory, most analytically inclined subjects start by cooperating but switch to a belligerent stance toward the end of the number of trials. But there is uncertainty where the switch will take place.

In Part A of a report I wrote in 1950 on the two-person non-zero-sum game for the ONR project, I considered the two-person Prisoner's Dilemma game repeated a fixed number of times (say 20). I assigned a *subjective probability* distribution over the way I thought others would play in order to figure out the best way I should play. In my naivete, without any theory or anything like that, I did what I now recognize as a prescriptive analysis for one party, making use of a descriptive modeling process for the possible decisions of the other parties. The descriptive modeling process involved assessing judgmental probability distributions. I slipped into being a subjectivist without realizing how radical I was behaving. That was the natural thing to do. No big deal.

Part B of the report dealt with complex non-zero-sum games where there's no solution. What would I do if two buddies of mine came over to me and said, "Look, Howard, we can't solve this game; there's no solution. You resolve it for us. What's fair?" Essentially I sought an arbitration rule that would propose a compromise solution for any non-zero-sum game. At that time I was familiar with the seminal work of Kenneth Arrow. Arrow sought a social welfare function that would combine individual preferences to arrive at a social or group preference. He examined a set of very plausible constraints on this social welfare function only to prove that these requirements were incompatible. No social welfare function exists that satisfies properties X, Y and Z. I adopted the Arrow approach: in my state of confusion, I proposed a set of reasonable desiderata for an arbitration scheme to satisfy and then I investigated their joint implications.

I remember distinctly how I started my research on arbitration rules. I attended a lecture by a labor arbitrator by the name of William Haber. During his talk about arbitration, I experienced an "aha" inspiration; I jumped out of my chair while the lecture was going on, I went back to my study and wrote vigorously for hours without a break on how I would arbitrate non-zero-sum games. That constituted Part B of my ONR report on non-zero-sum games. I used the Kakutani Fixed Point Theorem to show the existence of equilibria strategies.

That report was published informally in the Engineering Department. It was not peer reviewed; it was simply an informal report. At the time I was preparing to take my oral qualifying exam and searching for a thesis topic in linear, normed spaces, Banach spaces. This was in April of 1950. For the oral qualifying exams in the Math Department, the candidate first had to write a report on what he or she would like to be examined on; then, depending upon the report, the examiners structured the oral exam—its breadth and depth. In my written proposal I examined how all sorts of mathematical ideas found their way into the theory of stochastic processes. And then a surprising thing happened.

My wife, Estelle, received a telephone call from the very famous algebraist, Richard Brauer, who was the chairman of my oral examining committee. He informed her that, on the basis of my written report, the committee decided to excuse me from my oral exam. And then he said, "By the way, the committee would like to talk to me about my thesis." I came in the next day all excited about the fact that I didn't need to take my oral exam and was told that the committee thought it appropriate that I use my recently completed Engineering Report as my doctoral dissertation. I was stunned. So I ended up not having to take an oral exam, not having to write a thesis, and I was through before I thought I started.

**Fienberg**: So that was the end of your graduate education! Did you start immediately looking for a job?

**Raiffa:** That was in April; it was too late to go on to the job market and I didn't know what to do. The Departments of Mathematics and Psychology initiated an interdisciplinary seminar on Mathematics in the Social Sciences and as a post-doc I was hired to be the rapporteur of the seminar. I was charged to record what was said during the meetings and "what should have been said." I had a ball! For that one year I steeped myself mostly in psychological measurement theory, working with Clyde Coombs from psychology and Larry Klein from economics. That involvement constituted an important part of my analytical roots.

During my post-doc year I also gave a series of seminar talks to the statistical faculty and doctoral students on Abraham Wald's newly published book on *Statistical Decision Theory.* I was invited to do so because the book was very mathematical and made extensive use of game theory in existence proofs. Wald's book was full of Bayesian decision rules. Not as a way of making decisions but as a way of eliminating noncontenders—inadmissible rules. Wald never used subjective probabilities or judgments to choose a decision rule. Bayesian analysis was just a mathematical technique for finding out complete classes of admissible decision rules.

The following year I was ready to go on the job market and I had several offers from mathematics departments, but there were also two statistics ones: one from Columbia University's Department of Mathematical Statistics, the other one was working with George Shannon at Bell Labs. The Bell Labs job paid a lot more than Columbia. But Columbia presented a unique opportunity for me. A year earlier, Abraham Wald, who was the star of the statistics department at Columbia, was killed in an airplane accident over India and his colleagues Jack Wolfowitz and Jack Keifer left for Cornell when Wald died. The department was decimated but still there remained Ted Anderson, Henry Scheffe, Howard Levene and Herbert Solomon. But they needed someone desperately to teach Wald's stuff and to supervise his many doctoral students. I was supposed to fill that bill because I knew Wald's book.

I really was not prepared. Wald's doctoral students knew more statistics than I and there were times when I hadn't the faintest idea what they were talking about. I had to learn and give lectures on advanced topics in statistics. I did what I called "just-in-time" teaching. I used a book by Blackwell and Girshick. At that time it was available only in a pre-printed form and it hadn't yet been published. But it was a wonderful book. Blackwell and Girshick were more inclined toward the Bayesian viewpoint than Wald, but not really. (At least not then, although Blackwell later became an ardent supporter of the Bayesian perspective.) The book carefully skirted issues of measurability by confining itself to the denumerable case; it did nothing continuous. I bracketed their presentation by examining more closely the finite case, spending a lot of time on $n=2$, and then going abstract by considering the most advanced measure-theoretic version of the ideas. I produced copious notes for the course. I was a nervous wreck because my statistical colleagues all audited that course. I also taught the first course in statistics. I taught Neyman–Pearson theory, tests of hypotheses, confidence intervals, unbiased estimation and preached about the dangers of optional stopping that I no longer believe to be true.

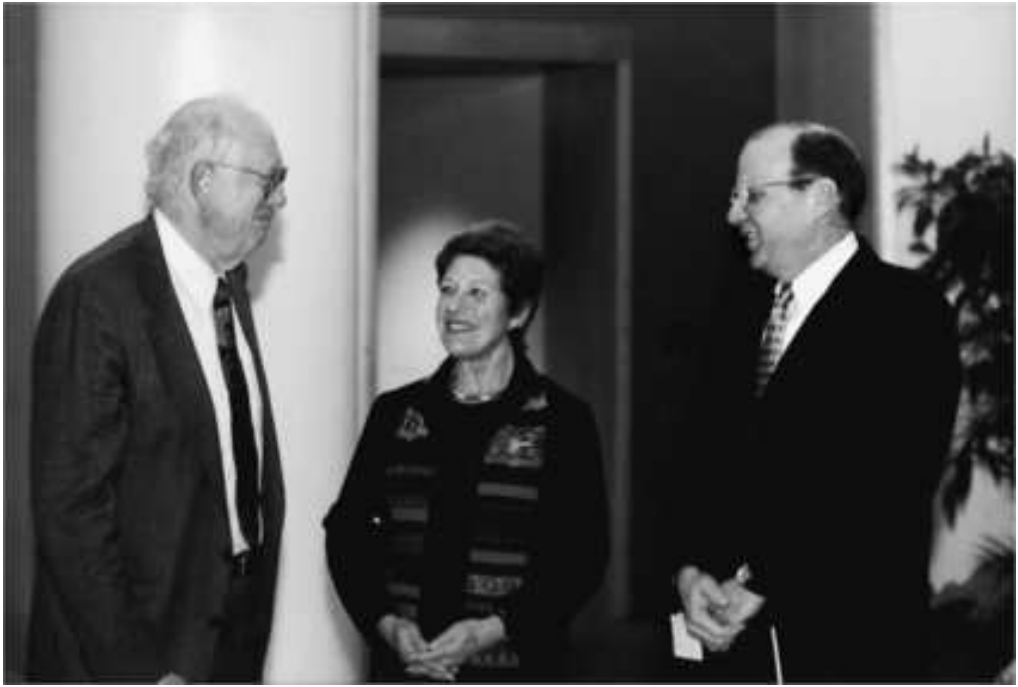

Fig. 2. *Raiffa with his wife Estelle and Carnegie Mellon President Jared L. Cohon at Dickson Prize Ceremony in April, 2000.*

Gradually I became disillusioned. I just didn't believe that the basic concepts I was teaching were central to what the field should be about. So I repeated what I knew how to do best. I went back to the Arrow approach and in an axiomatic style examined what primitive things that I believed in and explored their joint implications. At that time I should have known about some of the work of Jimmy Savage, but I didn't; I didn't even know who Jimmy Savage was. But in 1954 I came across a paper by Herman Chernoff justifying the Laplace solution to these problems. Essentially in a state of ignorance it argued for using a uniform prior distribution over states. I had my reservations about the full analysis, but Chernoff used an axiom he attributed to Herman Rubin, called the *Sure Thing Principle.* I embraced it wholeheartedly. The implications were devastating. It argued for making inferences and decisions based on the likelihood function. It also ruled out the Neyman–Pearson theory that preached that you can't infer what you should do based on an observed sample outcome until you think what you would do for all potential sample outcomes that could have occurred but didn't. "Nonsense," says the sure thing principle. Out went tests and unbiased estimates and confidence intervals and lots of stuff on optional stopping. Not much left. Now that we had destroyed so much, it was time to really seek an alternative to NP theory.

Consider the standard problem involving an unknown population proportion $p$. That's the usual binomial model. The sample produces nine $F$'s and one $S$ in that order and the likelihood function is $(1-p)^9 p^1$. Now compare that with the stopping rule that samples until the first success appears; and suppose that the first $S$ occurs on the tenth trial. The likelihood function would be the same whether you had the first stopping rule or the second stopping rule, and therefore the inference should be the same. But using Neyman–Pearson tests of hypothesis, the answers are different—you have to worry about not only what happened, but what could have happened according to the sampling plan.

I pushed the axiomatics and convinced myself that it made sense to assign a prior probability distribution over the states of the problem and maximize expected utility. It was for me akin to a religious conversion—from being a Neyman–Pearsonian to being a Bayesian. I became a closet Bayesian. I didn't come out of the closet because my associates, whom I admired, were vociferously opposed to Bayesianism. They thought it was a step backward. They'd say, "Look, Howard, what are you trying to do? Are you trying to introduce squishy judgmental, psychological stuff into something which we think is science?" Jimmy Savage, I think, had the best retort to this. He said: "Yes, I would rather build an edifice on the shifting sands of subjective probabilities than building on a void."

The biggest difference between me and my colleagues at Columbia was the kind of problems that we worked on. They were basically driven by the problem of inference. They paid lip service to decision problems by considering whether one should reject a null hypothesis, but really basically what they were interested in was problems of statistical sampling and inference going from observations to parameters. I came from a background in game theory and operational research, so for me a prototypical problem was: how much should you stock of a product when demand was unknown. For me, that unknown demand was the population value and that population value had a probability distribution. The whole point of doing sampling was to get better information, to get more informed probability distributions. Decisions were tied to real economic problems, not phony ones like should you accept a null hypothesis. That was really the divide, because my Columbia colleagues, whom I greatly admired, Henry Scheffé, Ted Anderson, Herb Robbins and others weren't interested in my class of problems. To me they were primarily inference people.

**Fienberg:** I notice from your resume that in 1957 you published the highly acclaimed book on *Games and Decisions* with Duncan Luce (Luce and Raiffa, 1957). You must have written this during the same period of time you were primarily learning and teaching and establishing your philosophical roots in statistics. Could you tell us about your game theory activities during this period?

**Raiffa:** A group of us started the interdisciplinary Behavioral Models Project at Columbia and we hired a mathematical psychologist, Duncan Luce, to supervise the project. Duncan was not affiliated with any department at the time. The principals were Paul Lazersfeld from psychology, Ernest Nagel from philosophy, Bill Vickery from economics, and I suppose I should include myself from statistics. As the junior member of that steering committee, I acted as Chairman of the project and Duncan and I carried the burden of pushing the project along. We had external financial support and ran seminars and hired pre-docs. We proposed publishing a series of short 50-page monographs on the topics that featured the use of mathematics in the social sciences—topics such as learning theory, psychological measurement theory, game theory, informatics and cybernetics—but we couldn't get any authors to submit manuscripts. So Duncan and I decided to write a 50-page document on games and decisions, on game theory really. We eventually gave up writing that fifty-page document and wrote instead a 500-page book on games and decisions. It took us two years to write. In 1953–1954, Duncan was at the Center for Advanced Study in the Behavioral Sciences at Stanford and I was at Columbia. The following year I went to that institute and he was back at Columbia. We were together face to face for five days in these two years, in an era well before e-mail, and we wrote the *Games and Decisions* book. It included a lot of material from my unpublished "non-dissertation."

**Fienberg:** That was a landmark book in many senses and it's still in print as a Dover paperback almost a half a century later. What happened next in your career?

**Raiffa:** The book still sells a few thousand copies a year. O.K., it is now '57 and out of the blue I got two offers: one from the University of California, Berkeley—not in statistics—and one from Harvard University—a joint appointment with the newly created Statistics Department and the Business School.

The newly formed Statistics Department was led by Fred Mosteller, a wonderful man and statistician, and it also included Bill Cochran, another great statistician. So I was very, very flattered, except I was worried. How would they receive my new conversion to Bayesianism? I talked to Fred Mosteller about that and he was lukewarm, but he was tolerant. And Cochran said: "Well, you'll grow up." At that time I literally was at Columbia for five years and I never knew that Columbia had a business school all this time. I really didn't know anything about business and the only reason I decided to go to Harvard was because of the Statistics Department. They were willing to double my Columbia salary. Columbia, belatedly, agreed to match it, and promote me, but we decided to go to Harvard. It was a close call in making that decision.

My wife, Estelle, and I stewed about the Harvard offer because there were many conflicting objectives we had to balance. We did a sort of formal analysis of this decision problem. Our analysis involved ten objectives that we scored and weighted. My wife is not mathematically inclined at all, but for this case she joined me in making all the assessments. It turned out that we agreed on practically everything. Harvard was the clear winner. Of course, there were some dimensions where Columbia was better, so it wasn't a dominating solution, but the formalization helped us really decide that it wasn't a close call at all. We then followed some advice that was given to us by Patty Lazersfeld. She advised that in decisions of this kind, don't ever make your choice without testing it. You tell your friends that you're going to Harvard, you tell your family, but you don't tell the administration. Then before you officially commit yourself, you see how you sleep for a week. And that's what we did. We slept well, we felt content, and we ended up at Harvard.

**Fienberg:** But when I arrived at Harvard in 1964 you were in essence full-time at the business school. How did this shift occur?

**Raiffa:** Surprisingly to me, my academic life didn't revolve around the statistics department; it revolved around a place called the Business School. At the B-School I worked closely with Robert O. Schlaifer. He's probably the person who influenced me more in my life than anybody else. He was trained as a classical historian and classical Greek scholar. During the war he worked for the underwater laboratory writing prose for technical reports. He ended up at the end of the war writing a tome on the engineering and economics of aviation engines. By some involvement, by some fluke, he received an appointment at the Business School but he had no specialty. The single professor at the B-School who taught a primitive course in statistics retired at that time and Robert was asked to teach that course. Thus history was made. Trained as a classical historian, he knew nothing about statistics, so he read the "classics": R. A. Fisher, Neyman and Pearson—not Wald and not Savage—and he concluded that standard statistical pedagogy did not address the main problem of a businessman: how to make decisions under uncertainty. Not knowing anything about the subjective/objective philosophical divide, he threw away the books and invented Bayesian decision theory from scratch. Since he had little mathematics, most of his examples involved discrete problems or the univariate case, like an unknown population proportion.

Robert had had only one course in mathematics, in the calculus. But he had raw mathematical abilities, provided he could see how it might be put to use. He was single-minded in his pursuit of relevance to the real world. When I came there, he was thrilled that here was a kindred soul that could tutor him in just the kind of mathematics he needed. I spent most of my days teaching Robert Schlaifer mathematics—first calculus, then linear algebra. I would teach him something about linear algebra in the morning and he would show me how it could be applied in the afternoon. He was not only smarter than other people, but he worked longer hours than anyone else.

**Fienberg:** I've heard you being alluded to as Mr. Decision Tree. What's the story behind that?

**Raiffa:** Already in my book with Luce on game theory, published in 1957, I used game trees to define games in extensive form. There were player nodes and chance nodes, but all chance nodes had associated, objective probabilities in the common knowledge domain. When I started working on individual decision problems at the B-School I modified the game tree into a decision tree that featured chance nodes with probability distributions subjectively assessed by the decision maker. My nonmathematically inclined audiences found it impossible to follow the logic of the analysis without an accompanying decision tree to keep track of the discussion. My use of decision trees started as a search for pedagogical simplicity, but I gradually became dependent on them myself. It's interesting to reflect why I never used decision trees earlier in teaching elementary statistics at Columbia. In *Applied Statistical Decision Theory* (ASDT), I present a schematic decision tree depicting the prototypical or canonical statistical decision problem. [At this point Raiffa went to the blackboard.]

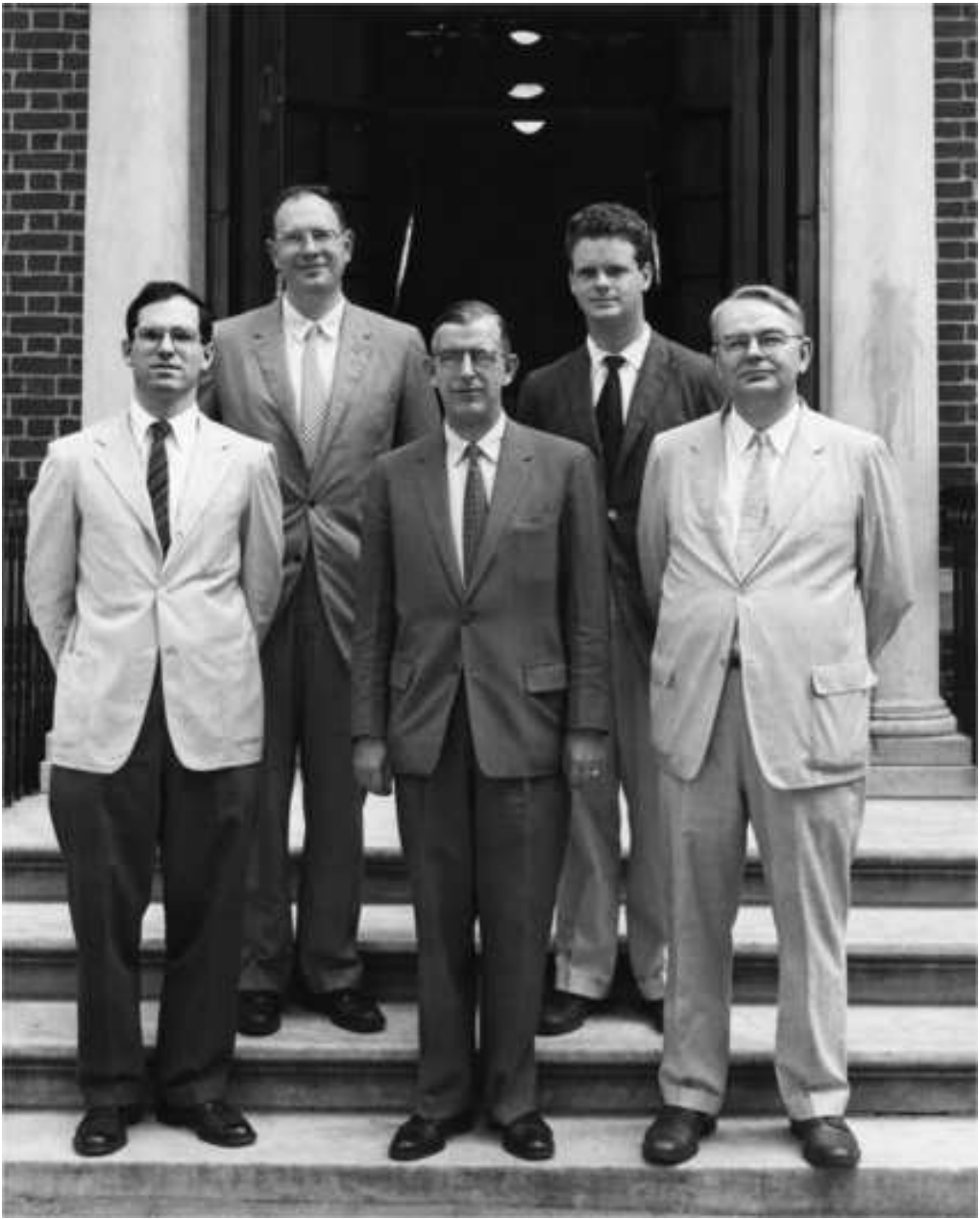

Fig. 3. *Harvard Statistics Department, 1957. From left to right: John Pratt, Raiffa, William Cochran, Arthur Dempster and Frederick Mosteller.*

At move 1, a decision node, the decision maker (DM) has a choice of experiments or information-gathering alternatives including the null experiment, which means acting now without gathering further information about the unknown population parameter, $\theta$. Move 2 is in chance's domain and the sample outcome is symbolically denoted by $z$. At move 3 the DM must choose a terminal act $a$ and at move 4 chance reveals the true population parameter $\theta$. This requires specifying the marginal probability of $z$ at move 2 and at move 4 the conditional or posterior probability of $\theta$ given $z$. To make these probability assessments, the DM usually starts with a subjectively assessed prior distribution of $\theta$ and an objective, model-based conditional sampling distri-bution of $z$ given each value of $\theta$. The DM then uses Bayes' theorem to find the probabilities required at move 4 and at 1. This is so standard that analysts using this methodology are called Bayesians. Classicists mistakenly choose not to assign probabilities at moves 2 and 4 because these involve subjective probability inputs and are taboo. Hence, for them, there is no gain to be had from considering this schematic decision tree.

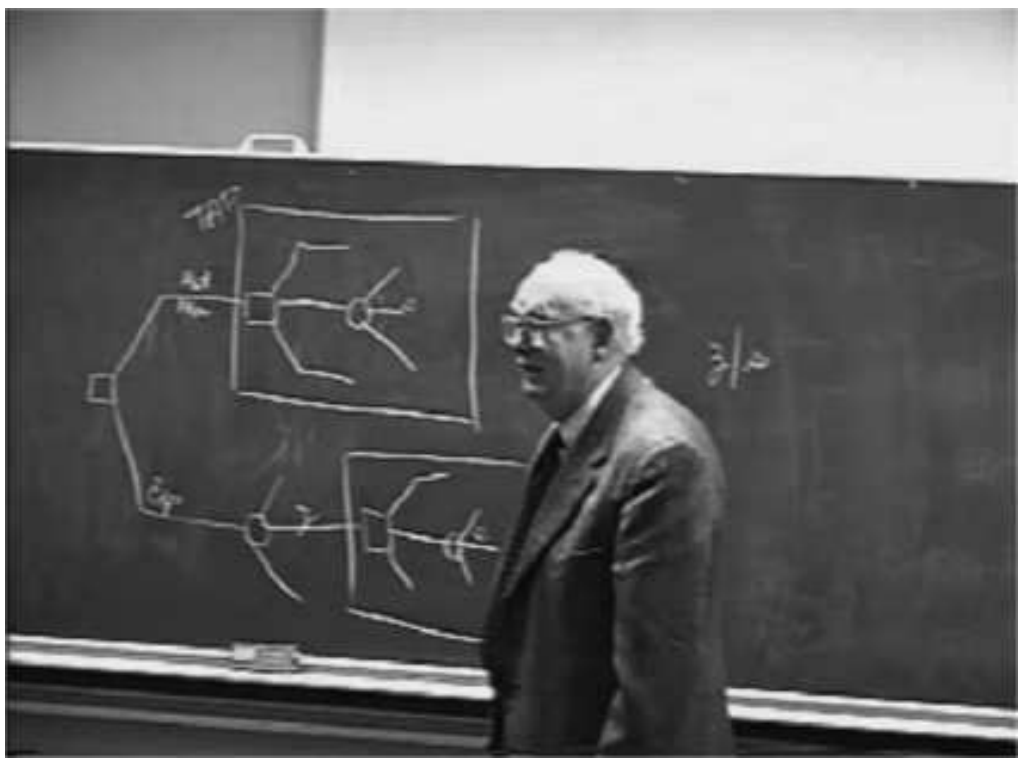

Fig. 4. *Raiffa drawing the decision tree.*

**Fienberg:** In 1957, when you really launched your efforts into the Bayesian direction, the word Bayesian was not used except perhaps in a pejorative or mathematical formalism kind of way. Jimmy Savage, Jack Good and Dennis Lindley did not call themselves Bayesians in the early parts of the fifties. Yet by the time you wrote the book with Bob, it had become second nature to identify yourself as such. How did that happen? How did Bayesian inference become known as "Bayesian"?

**Raiffa:** In the preface to ASDT we refer to "the so-called Bayesian approach." I'm not sure who is responsible for the nomenclature. But I dislike using the term "Bayesian" to refer to decision analysts who believe in using subjective probabilities because, once we generalize from the standard classical statistical paradigm, the schematic decision tree in the shown figure is too special and not indicative of the broader class of decision problems; and, in this wider class, subjective probability assignments are often made without invoking Bayes's formula.

**Fienberg:** So how would you like to be called?

**Raiffa:** I think of myself as a decision analyst who believes in using subjective probabilities. I would prefer being called a "subjectivist" than a "Bayesian." Robert and I divided Bayesians into two groups: engineers and scientists, or "*echt*" and the "non*echt*." The really true subjectivists were the engineers. The non*echt* scientists never elicited judgmental questions from anybody. For them it's all abstract. The *echt* folk got their hands dirty. In the early 1960s we had a series of distinguished Bayesians (Lindley, Box and Tiao), who each spent a semester at the Business School. They were primarily wonderful statisticians of the non-*echt* variety. The most notable of our *echt* visitors was Amos Tversky—but he visited us in the early 1980s. Next to Schlaifer, Amos was the person who influenced me most. But Amos was not a statistician.

**Fienberg:** Howard, why don't you continue with the story of your early interactions with Schlaifer.

**Raiffa:** Robert and I taught an elective course together in statistics at the B-School and it was a ball.

I opted for teaching both the objectivist and subjectivist points of view side by side with an openly declared preference for the subjective school. Robert thought that was a cop-out. He said, "Look, we're professors. We are supposed to know what's right. You teach what's right; you don't teach what's wrong." "But our students are going to have to read the literature," I retorted. "Since when do businessmen—never women—read the literature?" He wouldn't have anything to do with teaching Neyman–Pearson theory; it was judgmental probability all the way right down the line. The closest I could push him toward the classical school was to examine decision problems from the normal—as contrasted to the extensive—form of analysis. Not only was he smart, very smart, but he was the most opinionated person I ever met.

One objection we encountered in using the subjectivist approach to statistics was that it was too hard. Robert and I explored ways to simplify it. In 1959, after I was at the B-School two years, we started writing a book together on what you have alluded to as Bayesian statistics. It was not so much a book to be read or even a textbook but a compendium of results for the specialist. The theme of our book was: it doesn't have to be too complicated; anything the classicists can do we can do also—only better.

We discovered a simple algebraic way to go from priors to posteriors for sampling distributions that admitted fixed-dimensional sufficient statistics, like the exponential distributions.

**Kadane:** You mean the use of conjugate priors.

**Fienberg:** Well, you certainly succeeded in pushing the ideas. I'm opening the book right now, and here, almost at the beginning, there is a classical result on minimal sufficient statistics and exponential families. And then suddenly, as if out of nowhere, you introduce conjugate families and conjugate priors. Clearly, the ideas were around for special cases, going back at least into the nineteenth century, but I haven't found any other source that laid the approach out in full generality. How did you come to this idea?

**Raiffa:** Well, it was pretty obvious that if we're going to get a systematic Bayesian approach for the exponential-family distributions, we needed to get something where updating could be done algebraically in a formal sort of way. We needed the prior and posterior to belong to the same family of distributions and to conform well to the likelihood function. It's not hard to see how the mathematics would go. So I guess I can take credit for that.

**Fienberg:** You should!

**Raiffa:** I'll take responsibility for that one. It just seemed all so natural.

Our effort turned into the book you mentioned at the beginning, *Applied Statistical Decision Theory*, which I'm proud to say has been republished by Wiley in their classics series. Originally the book was published not by a regular publishing house but by the Harvard Business School Division of Research, which had never published anything mathematical before. The book must have sold maybe three hundred copies.

Jimmy Savage reviewed the manuscript very favorably and he called the notation *dazzlingly intricate*. He didn't like the notational conventions. I take responsibility for the intricate hieroglyphics. It works for me and seems to work also for novices who have not been brainwashed with usages of other notations. The theory is intricate enough so that when I'm away from the field for long periods of time and then return, I have a tough time remembering the theory and the notation comes to my rescue. Let me illustrate what I'm talking about. Let me go to the blackboard and show you. We consider a population parameter, designated by the Greek symbol $\mu$; since $\mu$ is an uncertain quantity, we flag it with a tilde sign, $\tilde{\mu}$. We distinguish between prior distributions and posterior distributions of $\tilde{\mu}$ by primes and double primes, giving respectively $\tilde{\mu}'$ and $\tilde{\mu}''$. We distinguish the prior mean—that is, the mean of $\tilde{\mu}'$—which we label $\bar{\mu}'$—from the posterior mean $\bar{\mu}''$. Now we get more complicated. From a prior point of view, we might be interested in the as-yet-unknown posterior mean. The distribution of this quantity was dubbed by Robert, the *pre-posterior* distribution and renamed by our students as the *preposterous* distribution. The prior variance of the as-yet-unknown posterior mean is connoted by $v'$. The posterior distribution of $\tilde{\mu}$ depends upon a sufficient statistic $z$, which we bring into our notational fold, and so it goes.

**Fienberg:** While that first printing of ASDT by the Business School may have not sold very many copies, a later paperback version brought out by MIT Press was widely used and important to those of us who tried to take the conjugate prior framework into statistical problems beyond decision theory. But then fairly soon after the book first appeared, you began to turn to related problems. How did this happen?

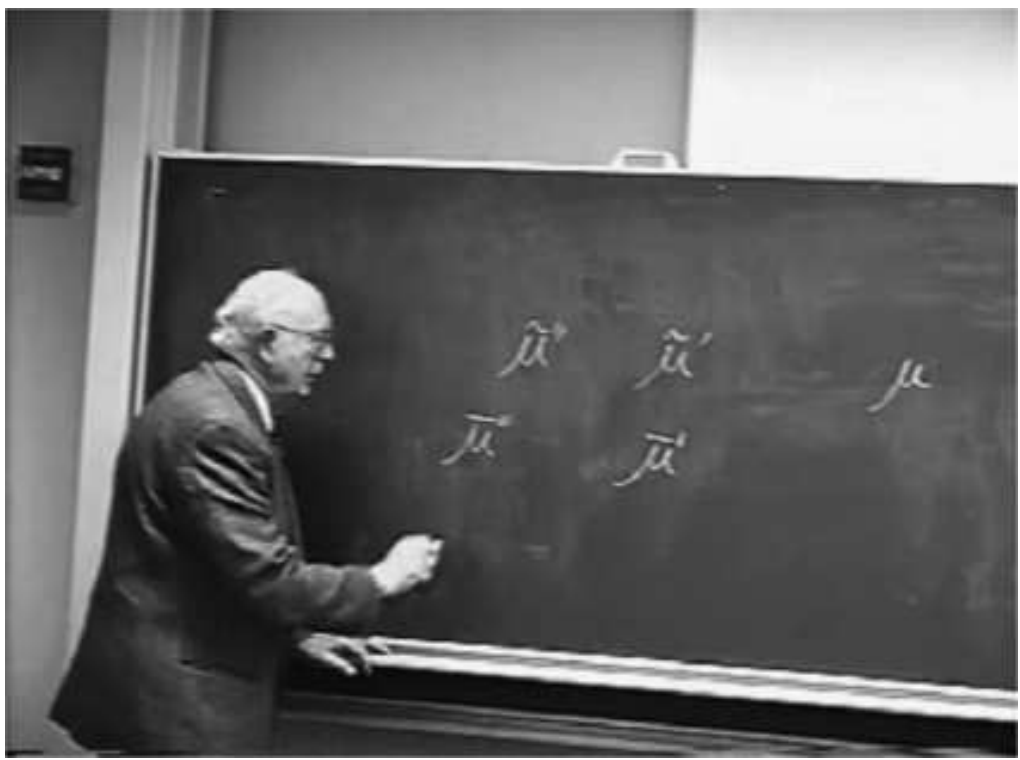

Fig. 5. *Raiffa recreating the ASDT notation at the blackboard.*

**Raiffa:** ASDT was written in 1959 and 1960 and published in '61. Schlaifer and I would discuss some ideas and my style was to start writing when I was still confused. The task of writing focused my mind. Schlaifer had trouble writing first drafts. He would look at what I had written and invariably his reaction was: "This is terrible," and he would tear up my version and write something better. But he had trouble starting without something to criticize.

In the academic year 1960–1961, I was awfully busy since I organized, at the request of the Ford Foundation, a special 11-month program for 40 research-oriented professors, teaching in management schools, who felt the need to learn more mathematics. That program, The Institute for Basic Mathematics for Application to Business (IBMAB), was reputed to be a huge success and the next generation of deans at such prestigious business schools as Harvard, Stanford and Northwestern were all graduates of that program. Naturally, in the IBMAB program I taught statistics from a subjectivist perspective with a heavy decision orientation and the gospel radiated outward in schools of management.

In the academic year 1961–1962, another radical thing happened to me. I was a second reader of a thesis proposal by Jack Grayson, a student in the Business School, in the area of finance. Grayson was interested in financial decisions of oil wildcatters. Through his interaction with me, his thesis was expanded from a purely descriptive to a prescriptive perspective. How should wildcatters accumulate information from geologic surveys, seismic soundings, exploratory wells, expert judgments? The drilling of an exploratory well was simultaneously a terminal-action and an information-gathering move. How should they form syndicates for the sharing of risks? This leads to decision problems galore, and the problems did not easily conform to the classical statistical decision paradigm involving sampling to gather information about an unknown population parameter. The statistical decision paradigm seemed too restrictive, too hobbling, too narrow. Schlaifer concurred with me and we began to think of ourselves more as decision analysts than as statisticians. I don't know why it took so long to make the shift, but in my mind every problem that I thought of up until that point was cast in the old statistical paradigm of going from a prior to a posterior distribution of a population parameter. With my new orientation I saw problems all over the place in business, in medicine, in engineering, in public policy where the decision problems under uncertainty did not fit comfortably into the classical mold. We were excited about our new vision of the world of uncertainty with a vast new agenda and we started the Decision under Uncertainty Seminar that you, Steve, referred to earlier.

**Fienberg:** This was also around the time that John Pratt worked with you and Robert on the introductory book that appeared in its preliminary edition in 1965. It was called *Introduction to Statistical Decision Theory*—ISDT in contrast to ASDT*. I studied from that unpublished manuscript. But then you never quite polished it up and finished it. What happened?

**Raiffa:** The reason why we didn't publish this massive 900-page book after we essentially finished it was that Schlaifer and I no longer believed in the centrality of the standard statistical paradigm as depicted in the above figure. The book was put aside because we had a new exciting agenda to explore. But we should have finished it. The fact that the book was available in a pre-publication form also took off the pressure for actually publishing it. That book wasn't finished until 1995, when I retired and had more time and more maturity. Pratt did most of the polishing and added more theoretical material, but Schlaifer chose not to get involved in the revisions.

We are now back in 1963 and with ISDT on a back burner, we eagerly pursued our new agenda. We had to learn how to elicit judgments from people—probabilities and utilities. Skeptics asserted that we couldn't get real experts to provide these subjective assessments. Well, we demonstrated to our satisfaction that was not the case. We worked with engineers and market experts who were more than willing to give us their subjective probability assessments for real problems. But were these assessments any good? Garbage in and garbage out. We realized that if we wanted to elicit judgments in a credible manner, there was the delicate issue on how you asked the questions. Framing was crucial. For example, we learned that if we asked questions about utility judgments in terms of incremental amounts, we would get different answers than if we asked people questions about their asset positions. And then we had to decide which set of responses should be used. Schlaifer and I convinced ourselves that questions should be posed in terms of asset positions and not in terms of incremental amounts of money, because increments invited all kinds of zero illusions.

I had a doctoral student who investigated the problem of overconfidence under my supervision. Experts did not calibrate very well; they were surprised a surprising number of times and we had to make our experts aware of this tendency. Our motivation was not description, but prescription, but nevertheless our students and we did a lot of work in what is now called behavioral decision making.

In the mid-1960s Robert introduced a required course in the first year of the MBA entitled Managerial Economics. All 800 students were exposed to cases that featured decision making under uncertainty. It was an heroic effort that was not universally appreciated by some of our nonnumerate students. But it was like an existence theorem: it demonstrated that decision analysis was relevant and teachable to future managers. In retrospect I think the effort was done too quickly without enough attention to palatable pedagogy. At semester's end the students burned one of Robert's books, one that I believe deserved a prize for innovation.

In the mid-1960s, I was offered and accepted a joint chair between the Business School and the Economics Department. I relished the fact that I never took a course in economics. I taught decision analysis, which in my mind had a different agenda than courses I once taught in statistical decision theory. I drifted away from Robert as I started to work on problems more in the public sector and to do research on multiple conflicting objectives and negotiations. But that research, together with my experiences at IIASA, are other chapters in my career which I'll discuss tomorrow at the Dickson Award ceremonies.

**Fienberg:** As you look back over the field of statistics, don't you have a sense of satisfaction about how your work with Robert and John has influenced others and the growth of the Bayesian school?

**Raiffa:** Certainly I'm proud of what I've contributed in this field. But still I'm a little disappointed. If we made a survey of the way statistics is taught across the country, it would be dominated by the old stuff, the Neyman–Pearson theory. Carnegie Mellon is a maverick; is an exception. The subjectivist school of decision making is not being taught in many places.

**Kass:** I think it's turned a corner.

**Raiffa:** Here, at Carnegie Mellon for sure, but...

**Kass:** Well, no, I think in the world, in the last ten years or so I think it's really started to change. And I don't think it's only in statistics but in lots of other fields, applied areas that really use statistical ideas and methods.

**Raiffa:** Well, let's look at the curricula of most of the universities. I think that overwhelmingly the classical school is still dominant at most universities. I was instrumental in helping to start the Kennedy School of Government at Harvard and in the beginning I had some input in what was taught. In the early days the curriculum was decision and policy oriented and Bayesian statistics and decision analysis were taught and integrated in the curriculum. But new teachers were hired and they taught what they had learned as students and Bayesianism disappeared.

**Kass:** I have another question; it's almost the same as Steve's, involving the Bayes part of it. One of the things that's so interesting to many of us is to examine the way one body of work influences another. And in the case of your book with Schlaifer, it's easy to see how that influenced, for instance, Morrie DeGroot's book, which came a decade or so later, and then, much later, Jim Berger's book which most students today are familiar with in modern statistical decision theory. In retrospect it's clear how your book inspired the material and presentation in DeGroot's book. To see these three books, one right after another, it's very easy to trace the influences backwards, but how do we trace things back to see the influences on your work with Robert as a Bayesian?

**Raiffa:** Schlaifer was driven by the need to coordinate statistics with business decision making and he truly discovered from scratch the basic ideas of what you refer as Bayesianism. I, on the other hand, was brainwashed into the classical tradition and had to go through a religious conversion.

**Kadane:** What about Jimmy Savage's work and his 1954 book?

**Raiffa:** Somehow I just was not aware of that book until I left Columbia. I already mentioned Herman Chernoff's paper and Herman Rubin's sure thing principle. That had a profound effect on me.

**Kass:** Not only has the Bayesian world become more intimately involved in applications since your early efforts, but statistics as a whole has moved in this direction.

**Raiffa:** I hope you are correct but it's painfully slow. I look forward to the day that there will be Departments of Decision Sciences in other universities besides Carnegie-Mellon and Duke. Statistics is a broad subject encompassing data analysis, modeling and inference as well as decisions. I just don't want the decision component to disappear.

**Fienberg:** Well, Howard, we all look forward to a continuation of this conversation when you can tell us more about your analytical roots as a decision scientist and about your experiences after 1967.

**Raiffa:** I look forward to it.

## REFERENCES

HAMMOND, J. S., KEENEY, R. L. and RAIFFA, H. (1998). *Smart Choices.* Harvard Business School Press, Boston.

KEENEY, R. L. and RAIFFA, H. (1976). *Decisions with Multiple Objectives*: *Preferences and Value Tradeoffs.* Wiley, New York. Reprinted, Cambridge Univ. Press, New York (1993). MR0449476

LUCE, R. D. and RAIFFA, H. (1957). *Games and Decisions*: *Introduction and Critical Survey.* Wiley, New York. Paperback reprint, Dover, New York. MR0087572

PRATT, J. W., RAIFFA, H. and SCHAIFER, R. (1995). *Introduction to Statistical Decision Theory.* MIT Press, Cambridge, MA. MR1326829

RAIFFA, H. (1968). *Decision Analysis*: *Introductory Lectures on Choices Under Uncertainty.* Addison-Wesley, Reading, MA.

RAIFFA, H. (1982). *The Art and Science of Negotiation.* Harvard Univ. Press, Cambridge, MA.

RAIFFA, H. (2002). *Negotiation Analysis.* Harvard Univ. Press, Cambridge, MA.

RAIFFA, H., RICHARDSON, J. and METCALFE, D. (2003). *Negotiation Analysis*: *The Science and Art of Collaborative Decision.* Harvard Univ. Press, Cambridge, MA.

RAIFFA, H. and SCHAIFER, R. (1961). *Applied Statistical Decision Theory.* Division of Research, Harvard Business School, Boston. 1968 paperback edition, MIT Press, Cambridge, MA. Wiley Classics Library edition (2000).